# **On the existence of negative energy states in QED in the temporal gauge**

by

Dan Solomon

Rauland-Borg Corporation

3450 W. Oakton

Skokie, IL USA

Email: dan.solomon@rauland.com

## Abstract

It is generally assumed that the vacuum state is the quantum state with the lowest energy. However, it has been shown that this is not the case for a Dirac-Maxwell field in the temporal gauge. In this paper we will present another proof, different from that presented in previous work, which shows that the vacuum state is not the minimum energy state for a Dirac-Maxwell field in the temporal gauge.

## 1. Introduction.

It is generally assumed in quantum mechanics that the vacuum state is the state of minimum energy. However recent work has shown that this assumption is not necessarily valid. For example, it has been shown in Dirac hole theory that there exist quantum states with less energy than the vacuum state [1,2]. It has also been shown that, in Dirac field theory, the assumption that the vacuum is the state of minimum energy is in conflict with the requirement of gauge invariance and that the theory will be mathematically inconsistent if the vacuum is required to be the state of minimum energy [2,3,4]. In addition, it has also been shown that the vacuum state is not a minimum energy state for a Dirac-Maxwell field in the temporal gauge [5].

In light of this research the question of whether or not the vacuum state is the state of minimum energy should not be assumed but should be carefully examined and proved or disproved for the quantum theory in question. Therefore, in this paper, we shall again examine a Dirac-Maxwell field in the temporal gauge and show that the vacuum is not the minimum energy state. However the proof presented here will be different from that of previous work. Note that throughout this discussion we will use $\hbar = c = 1$. Also vectors are indicated by bold text.

## 2. QED in the temporal gauge.

If $|\Omega\rangle$ is a normalized state vector and $\hat{H}$ is the Hamiltonian then the energy is given by $\langle\Omega|\hat{H}|\Omega\rangle$. The question we want to examine is whether or not there exists a lower bound to the energy of a quantum state for a Dirac-Maxwell field in the temporal gauge. That is, does there exist a normalized state vector $|\Omega_{vac}\rangle$, usually considered to the be the vacuum state, where,

$$\langle\Omega|\hat{H}|\Omega\rangle \geq \langle\Omega_{vac}|\hat{H}|\Omega_{vac}\rangle \text{ for all normalized state vectors } |\Omega\rangle \tag{2.1}$$

The Dirac-Maxwell field describes the interaction between a quantized fermion field and quantized electromagnetic field. In the temporal gauge the gauge condition is given by the relationship $A_0 = 0$ [5,6,7,8,9] where $A_0$ is the scalar component of the electric potential. The Hamiltonian $\hat{H}$ is given by [5,8],

$$\hat{H} = \hat{H}_{0,D} + \hat{H}_{0,M} - \int \hat{\mathbf{J}}(\mathbf{x}) \cdot \hat{\mathbf{A}}(\mathbf{x}) d\mathbf{x} - \varepsilon_R \tag{2.2}$$

The quantities in the above expression are defined by,

$$\hat{H}_{0,D} = \frac{1}{2} \int \left[ \hat{\psi}^\dagger(\mathbf{x}), H_{0,D} \hat{\psi}(\mathbf{x}) \right] d\mathbf{x}; \quad H_{0,D} = -i\boldsymbol{\alpha} \cdot \nabla + \beta m \tag{2.3}$$

$$\hat{H}_{0,M} = \frac{1}{2} \int \left( \hat{\mathbf{E}}^2 + \hat{\mathbf{B}}^2 \right) d\mathbf{x}; \qquad \hat{\mathbf{B}}(\mathbf{x}) = \nabla \times \hat{\mathbf{A}}(\mathbf{x}) \tag{2.4}$$

$$\hat{\mathbf{J}}(\mathbf{x}) = \frac{q}{2} \left[ \hat{\psi}^\dagger(\mathbf{x}), \boldsymbol{\alpha} \hat{\psi}(\mathbf{x}) \right] \tag{2.5}$$

In the above expressions $m$ is the fermion mass, $\boldsymbol{\alpha}$ and $\beta$ are the usual 4x4 matrices, $q$ is the electric charge, $\hat{H}_{0,D}$ is the Dirac Hamiltonian, $\hat{H}_{0,M}$ is the Hamiltonian for the electromagnetic field, $\hat{\mathbf{J}}(\mathbf{x})$ is the current operator, and $\varepsilon_R$ is a renormilazation constant which is chosen so that the energy of the vacuum state will be zero. The fermion field operators are $\hat{\psi}(\mathbf{x})$ and $\hat{\psi}^\dagger(\mathbf{x})$ and the field operators for the electromagnetic field are $\hat{\mathbf{A}}(\mathbf{x})$ and $\hat{\mathbf{E}}(\mathbf{x})$. The electromagnetic field operators are real so that $\hat{\mathbf{A}}^\dagger(\mathbf{x}) = \hat{\mathbf{A}}(\mathbf{x})$ and $\hat{\mathbf{E}}^\dagger(\mathbf{x}) = \hat{\mathbf{E}}(\mathbf{x})$.

The field operators obey the following relationships,

$$\left[ \hat{A}^i(\mathbf{x}), \hat{E}^j(\mathbf{y}) \right] = -i\delta_{ij}\delta^3(\mathbf{x} - \mathbf{y}); \left[ \hat{A}^i(\mathbf{x}), \hat{A}^j(\mathbf{y}) \right] = \left[ \hat{E}^i(\mathbf{x}), \hat{E}^j(\mathbf{y}) \right] = 0 \tag{2.6}$$

and

$$\left\{ \hat{\psi}_a^\dagger(\mathbf{x}), \hat{\psi}_b(\mathbf{y}) \right\} = \delta_{ab}\delta(\mathbf{x} - \mathbf{y}); \left\{ \hat{\psi}_a^\dagger(\mathbf{x}), \hat{\psi}_b^\dagger(\mathbf{y}) \right\} = \left\{ \hat{\psi}_a(\mathbf{x}), \hat{\psi}_b(\mathbf{y}) \right\} = 0 \tag{2.7}$$

where "a" and "b" are spinor indices. In addition, all commutators between the electromagnetic field operators and fermion field operators are zero, i.e.,

$$\left[ \hat{\mathbf{A}}(\mathbf{x}), \hat{\psi}(\mathbf{y}) \right] = \left[ \hat{\mathbf{E}}(\mathbf{x}), \hat{\psi}(\mathbf{y}) \right] = \left[ \hat{\mathbf{A}}(\mathbf{x}), \hat{\psi}^\dagger(\mathbf{y}) \right] = \left[ \hat{\mathbf{E}}(\mathbf{x}), \hat{\psi}^\dagger(\mathbf{y}) \right] = 0 \tag{2.8}$$

Define the operator,

$$\hat{G}(\mathbf{x}) = \nabla \cdot \hat{\mathbf{E}}(\mathbf{x}) - \hat{\rho}(\mathbf{x}) \tag{2.9}$$

where the current operator $\hat{\rho}(\mathbf{x})$ is defined by,

$$\hat{\rho}(\mathbf{x}) = \frac{q}{2} \left[ \hat{\psi}^\dagger(\mathbf{x}), \hat{\psi}(\mathbf{x}) \right] \tag{2.10}$$

All physically acceptable state vectors $\left|\Omega\right\rangle$ must satisfy the gauss's law constraint,

$$\hat{G}(\mathbf{x})\left|\Omega\right\rangle = 0 \tag{2.11}$$

The expression for the Hamiltonian along with the commutator relationships can be used to obtain the operator form of the equations of motion. These include Maxwell's equations and the continuity equation. In classical physics the continuity equation is,

$$\frac{\partial \rho}{\partial t} + \nabla \cdot \mathbf{J} = 0 \tag{2.12}$$

In quantum mechanics the time derivative $\partial\rho/\partial t$ is replaced by $i\left[\hat{H},\hat{\rho}\right]$ to obtain,

$$\left[\hat{H},\hat{\rho}\right] = i\nabla \cdot \hat{\mathbf{J}} \tag{2.13}$$

**3. Negative energy states.**

The vacuum state $\left|\Omega_{vac}\right\rangle$ must satisfy Eq. (2.11), i.e.,

$$\hat{G}(\mathbf{x})\left|\Omega_{vac}\right\rangle = 0 \tag{3.1}$$

It is an eigenstate of the Hamiltonian with an eigenvalue of zero,

$$\hat{H}\left|\Omega_{vac}\right\rangle = 0 \tag{3.2}$$

Note the eigenstate is zero due to proper selection of the renormalization constant $\varepsilon_R$. This is done in order to simplify the following discussion. Obviously, then, the energy of the vacuum state $\left\langle\Omega_{vac}\right|\hat{H}\left|\Omega_{vac}\right\rangle = 0$. Now we want to determine if there exist any states $\left|\Omega\right\rangle$ for which the energy $\left\langle\Omega\right|\hat{H}\left|\Omega\right\rangle$ is less than zero. Define the operator,

$$\hat{O}_a = \exp\left(i\int \chi(\mathbf{x})G(\mathbf{x})d^3x\right) \tag{3.3}$$

where $\chi(\mathbf{x})$ is an arbitrary real valued function. Use the fact that $\hat{\rho}(\mathbf{x})$ commutes with $\hat{\mathbf{E}}(\mathbf{x})$ to obtain,

$$\hat{O}_a = \hat{O}_2\hat{O}_1 \tag{3.4}$$

where,

$$O_1 = \exp\left(i\int \chi(\mathbf{x})\nabla\cdot\mathbf{E}(\mathbf{x})d^3x\right) = \exp\left(-i\int \mathbf{E}(\mathbf{x})\cdot\nabla\chi(\mathbf{x})d^3x\right) \tag{3.5}$$

and,

$$O_2 = \exp\left(-i\int \chi(\mathbf{x})\hat{\rho}(\mathbf{x})d^3x\right) \tag{3.6}$$

where integration by parts has been used in deriving (3.5). We can define new states by acting on existing states with an operator [10]. Define the states,

$$|\Omega_1\rangle = O_1|\Omega_{vac}\rangle \text{ and } |\Omega_2\rangle = O_2|\Omega_{vac}\rangle \tag{3.7}$$

It is easy to show that $\left[\hat{O}_1, \hat{G}(\mathbf{x})\right] = \left[\hat{O}_2, \hat{G}(\mathbf{x})\right] = 0$, therefore $|\Omega_1\rangle$ and $|\Omega_2\rangle$ satisfy (2.11) so that $|\Omega_1\rangle$ and $|\Omega_2\rangle$ are physically acceptable states.

Now, using the fact the vacuum state $|\Omega_{vac}\rangle$ satisfies (3.1), we have that,

$$O_a|\Omega_{vac}\rangle = |\Omega_{vac}\rangle \tag{3.8}$$

Using this along with (3.2) and (3.4) we obtain,

$$\langle\Omega_{vac}|O_a^\dagger \hat{H} O_a|\Omega_{vac}\rangle = \langle\Omega_{vac}|O_1^\dagger O_2^\dagger \hat{H} O_2 O_1|\Omega_{vac}\rangle = 0 \tag{3.9}$$

Next we will use the Baker-Campell-Hausdorff relationships which state that,

$$e^{+\hat{O}_A}\hat{O}_B e^{-\hat{O}_A} = \hat{O}_B + \left[\hat{O}_A, \hat{O}_B\right] + \frac{1}{2}\left[\hat{O}_A, \left[\hat{O}_A, \hat{O}_B\right]\right] + \ldots \tag{3.10}$$

where $\hat{O}_A$ and $\hat{O}_B$ are arbitrary operators. From this and (3.6) we obtain,

$$O_2^\dagger \hat{H} O_2 = \hat{H} + i\left[\hat{C}, \hat{H}\right] + \frac{i^2}{2}\left[\hat{C}, \left[\hat{C}, \hat{H}\right]\right] + \ldots \tag{3.11}$$

where,

$$\hat{C} = \int \chi(\mathbf{x})\hat{\rho}(\mathbf{x})d^3x \tag{3.12}$$

Use (2.13) to obtain,

$$\left[\hat{C}, \hat{H}\right] = -i\int \chi\nabla\cdot\mathbf{J}d^3x = i\int \mathbf{J}\cdot\nabla\chi d^3x \tag{3.13}$$

where we have done an integration by parts. Use this in (3.11) to obtain,

$$O_2^\dagger \hat{H} O_2 = \hat{H} + \hat{D} \tag{3.14}$$

where,

$$\hat{D} \equiv -\int \mathbf{J}\cdot\nabla\chi d^3x + \frac{i^2}{2}\left[\hat{C}, i\int \mathbf{J}\cdot\nabla\chi d^3x\right] + \ldots \tag{3.15}$$

Use this result and (3.2) to obtain,

$$\langle\Omega_{vac}|O_2^\dagger \hat{H} O_2|\Omega_{vac}\rangle = \langle\Omega_{vac}|\hat{H}|\Omega_{vac}\rangle + \langle\Omega_{vac}|\hat{D}|\Omega_{vac}\rangle = \langle\Omega_{vac}|\hat{D}|\Omega_{vac}\rangle \tag{3.16}$$

Now $\hat{D}$ is a function of fermion operators only. Therefore $\left[\hat{O}_1, \hat{D}\right] = 0$ due to the fact that $\hat{\mathbf{E}}$ commutes with all fermion operators. Use this fact along with $O_1^\dagger O_1 = 1$ and (3.14) to obtain,

$$O_1^\dagger O_2^\dagger \hat{H} O_2 O_1 = O_1^\dagger \left(\hat{H} + \hat{D}\right) O_1 = O_1^\dagger \hat{H} O_1 + \hat{D} O_1^\dagger O_1 = O_1^\dagger \hat{H} O_1 + \hat{D} \tag{3.17}$$

Use (3.9) and the above to obtain,

$$\left\langle \Omega_{vac} \right| \left( O_1^\dagger \hat{H} O_1 + \hat{D} \right) \left| \Omega_{vac} \right\rangle = 0 \tag{3.18}$$

This yields.

$$\left\langle \Omega_{vac} \right| O_1^\dagger \hat{H} O_1 \left| \Omega_{vac} \right\rangle = -\left\langle \Omega_{vac} \right| \hat{D} \left| \Omega_{vac} \right\rangle \tag{3.19}$$

Now from this result and (3.16) we obtain,

$$\left\langle \Omega_{vac} \right| O_2^\dagger \hat{H} O_2 \left| \Omega_{vac} \right\rangle = -\left\langle \Omega_{vac} \right| O_1^\dagger \hat{H} O_1 \left| \Omega_{vac} \right\rangle \tag{3.20}$$

Next use this result and (3.7) to obtain,

$$\left\langle \Omega_1 \right| \hat{H} \left| \Omega_1 \right\rangle = -\left\langle \Omega_2 \right| \hat{H} \left| \Omega_2 \right\rangle \tag{3.21}$$

Recall that the energy of the vacuum state $\left\langle \Omega_{vac} \right| \hat{H} \left| \Omega_{vac} \right\rangle = 0$. Therefore if the energy of $\left| \Omega_1 \right\rangle$ is positive then the energy of $\left| \Omega_2 \right\rangle$ is negative and vica-versa. Therefore the only way to satisfy condition (2.1) as well as (3.21) is to have,

$$\left\langle \Omega_1 \right| \hat{H} \left| \Omega_1 \right\rangle = \left\langle \Omega_2 \right| \hat{H} \left| \Omega_2 \right\rangle = 0 \tag{3.22}$$

Suppose that this is the case. Now consider the state $\left| \Omega_2 \right\rangle$. Now even though $\left\langle \Omega_2 \right| \hat{H} \left| \Omega_2 \right\rangle = 0$ it can be shown that $\hat{H} \left| \Omega_2 \right\rangle \neq 0$. This is done in the appendix. Therefore it is reasonable to assume that there exists another physically acceptable state $\left| \Omega' \right\rangle$ such that,

$$\left\langle \Omega' \right| \hat{H} \left| \Omega_2 \right\rangle \neq 0 \tag{3.23}$$

Now define the state,

$$\left| \Omega'' \right\rangle = N\left( \left| \Omega_2 \right\rangle + \alpha \left| \Omega' \right\rangle \right) \tag{3.24}$$

where $N$ is a normalization constant so that $\left\langle \Omega'' \middle| \Omega'' \right\rangle = 1$. $\left| \Omega'' \right\rangle$ satisfies (2.11), therefore it is a physically acceptable state. The energy of $\left| \Omega'' \right\rangle$ is given by,

$$\langle\Omega''|\hat{H}|\Omega''\rangle = N^2\left(\langle\Omega_2| + \alpha^*\langle\Omega'|\right)\hat{H}\left(|\Omega_2\rangle + \alpha|\Omega'\rangle\right) \tag{3.25}$$

This yields,

$$\langle\Omega''|\hat{H}|\Omega''\rangle = N^2\left(\langle\Omega_2|\hat{H}|\Omega_2\rangle + \alpha^*\langle\Omega'|\hat{H}|\Omega_2\rangle + \alpha\langle\Omega_2|\hat{H}|\Omega'\rangle + |\alpha|^2\langle\Omega'|\hat{H}|\Omega'\rangle\right) \tag{3.26}$$

Use (3.22) to obtain,

$$\langle\Omega''|\hat{H}|\Omega''\rangle = N^2\left(\alpha^*\langle\Omega'|\hat{H}|\Omega_2\rangle + \alpha\langle\Omega_2|\hat{H}|\Omega'\rangle + |\alpha|^2\langle\Omega'|\hat{H}|\Omega'\rangle\right) \tag{3.27}$$

Now in the limit $\alpha \to 0$ we can drop the $|\alpha|^2$ term to obtain,

$$\langle\Omega''|\hat{H}|\Omega''\rangle \underset{\alpha\to 0}{=} N^2\left(\alpha^*\langle\Omega'|\hat{H}|\Omega_2\rangle + \alpha\langle\Omega_2|\hat{H}|\Omega'\rangle\right)$$

Now since $\langle\Omega'|\hat{H}|\Omega_2\rangle \neq 0$ it is evident we can find an $\alpha$ such that $\langle\Omega''|\hat{H}|\Omega''\rangle < 0$ so that the energy of the state $|\Omega''\rangle$ will be less than that of the vacuum state $|\Omega_{vac}\rangle$. Therefore we have shown that, for a Dirac-Maxwell field in the temporal gauge, there exist physically acceptable quantum states with less energy than that of the vacuum state.

**Appendix**

We want to prove that $\hat{H}|\Omega_2\rangle \neq 0$. First we write,

$$\hat{H}|\Omega_2\rangle = \hat{H}\hat{O}_2|\Omega_{vac}\rangle = \left(\left[\hat{H},\hat{O}_2\right] + \hat{O}_2\hat{H}\right)|\Omega_{vac}\rangle = \left[\hat{H},\hat{O}_2\right]|\Omega_{vac}\rangle$$

Now,

$$\hat{O}_2 = e^{-i\hat{C}} = 1 - i\hat{C} + \frac{(-i)^2}{2}\hat{C}^2 + \ldots$$

Due to the fact that $\chi(\mathbf{x})$ is an arbitrary function in order for $\left[\hat{H},\hat{O}_2\right]|\Omega_{vac}\rangle = 0$ for all $\chi(\mathbf{x})$ we must have that $\left[\hat{H},\hat{\rho}(\mathbf{x})\right]|\Omega_{vac}\rangle = 0$. Using (2.13) this would imply that $\nabla\cdot\hat{\mathbf{J}}(\mathbf{x})|\Omega_{vac}\rangle = 0$. However there is a theorem that states if a local operator $\hat{T}(\mathbf{x})|\Omega_{vac}\rangle = 0$ then $\hat{T}(\mathbf{x}) = 0$ (See theorem 4-3 of Streater and Wightmann[11] or Epstein et al[12]). Now $\nabla\cdot\hat{\mathbf{J}}(\mathbf{x})$ is local operator and $\nabla\cdot\mathbf{J}(\mathbf{x}) \neq 0$ therefore $\nabla\cdot\mathbf{J}(\mathbf{x})|\Omega_{vac}\rangle \neq 0$. Therefore we have that $\hat{H}|\Omega_2\rangle \neq 0$.